\ifdefined\XeTeXversion\else\pdfoutput=1\fi
\documentclass[conference]{IEEEtran}
\usepackage[T1]{fontenc}
\usepackage{newtxtext}
\usepackage{newtxmath}
\usepackage[scaled=0.92]{helvet}
\ifdefined{texgyreheros}[Extension=.otf,UprightFont=*-regular,BoldFont=*-bold,ItalicFont=*-italic,BoldItalicFont=*-bolditalic,Scale=0.92]\fi
\usepackage[final]{microtype}
\usepackage[nocompress]{cite}
\usepackage{amsmath}
\usepackage{booktabs}
\usepackage{tabularx}
\usepackage{array}
\usepackage{url}
\usepackage{graphicx}
\usepackage{tikz}
\usetikzlibrary{positioning,arrows.meta,fit,backgrounds}
\usepackage{hyperref}
\definecolor{osLink}{RGB}{0,92,138}
\hypersetup{colorlinks=true,linkcolor=osLink,citecolor=osLink,urlcolor=osLink,pdftitle={Superintelligent Software Engineering: Governance That Scales with Capability},pdfauthor={Mohammad Mehdi Gholinejad}}%%
\usepackage{balance}
\makeatletter
\def\abstract{\normalfont\@IEEEabskeysecsize\bfseries\textit{\abstractname}:\enspace\relax
  \@IEEEgobbleleadPARNLSP}
\def\IEEEkeywords{\normalfont\@IEEEabskeysecsize\bfseries\textit{\IEEEkeywordsname}:\enspace\relax
  \@IEEEgobbleleadPARNLSP}
\makeatother

\newcommand{\op}{\textit{[Opinion]}}
\newcommand{\exper}{\textit{[Experience]}}
\newcommand{\bone}{\textsc{bone}}
\newcolumntype{Y}{>{\raggedright\arraybackslash}X}
\newenvironment{filelist}{\begin{itemize}\raggedright}{\end{itemize}}

\begin{document}

\title{Superintelligent Software Engineering:\\Governance That Scales with Capability}

\author{\IEEEauthorblockN{Mohammad Mehdi Gholinejad}
\IEEEauthorblockA{OSTEON Labs\\
ORCID 0009-0004-4192-9825}}

\maketitle

\begin{abstract}
Software agents are being trained to exceed human engineers; we argue that capability alone does
not produce better software engineering. We define
\emph{superintelligent software engineering} as the discipline of building and evolving software
with agents whose capability exceeds that of human engineers, in which governance (gates,
contracts, decisions and decay control) scales with capability. The argument rests on three claims.
First, correctness in a project rests on decisions and contracts that the code does not carry.
Second, knowledge decays. Third, once an agent outpaces its reviewers, reading catches a shrinking
share of the changes built on stale knowledge. A simple model shows that, past
a fixed review capacity, escaping changes grow linearly with throughput while the share that
reading catches falls toward zero; governance can lower the slope, capability alone cannot. An
exploratory snapshot of 452 of the most-starred public GitHub repositories finds a root-level or %%
\texttt{.github} agent context file in 48.67\% (95\% confidence interval 44.1 to 53.3); in the %%
median file that names repository paths, 9.5\% to 22\% of those paths do not resolve, depending on %%
how a reference is read. We then present OSTEON, a discipline for the regime, archived on Zenodo
under DOI \mbox{10.5281/zenodo.23436020} as a plugin for Claude Code. In OSTEON, each project gets a governed knowledge repository of decisions and
contracts, every change is checked mechanically against it, and the knowledge itself is checked for
decay at every session start; planning is gated, and sessions record what they learn without
writing rules. We report no measured effect of OSTEON, mark claims about its mechanisms as
opinion and the author's practice heuristics as design rationale, and state the questions that
would falsify the thesis.
\end{abstract}


\section{Introduction}

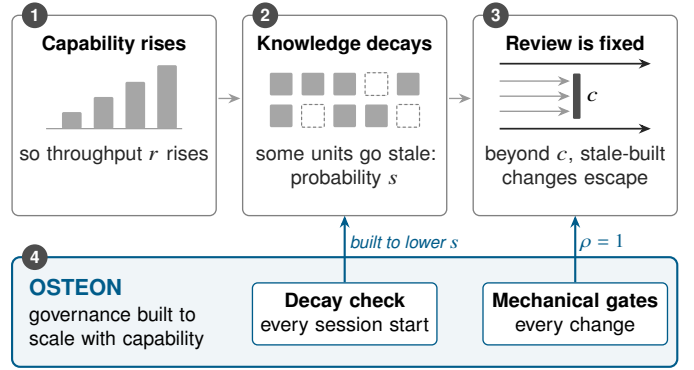
\begin{figure}[!t]
\centering
{\sffamily\ifdefined\fontspec\else\fontencoding{T1}\fontfamily{phv}\selectfont\fi
\providecolor{osAccent}{RGB}{0,92,138}%%
\begin{tikzpicture}[
  x=1cm, y=1cm,
  font=\footnotesize,
  panel/.style={draw=black!50, line width=0.5pt, rounded corners=2.5pt, fill=white},
  gov/.style={draw=osAccent, line width=0.8pt, rounded corners=3pt, fill=osAccent!6},
  mech/.style={draw=osAccent, line width=0.5pt, rounded corners=2pt, fill=white,
    align=center, text width=2.3cm, inner sep=2pt, minimum height=0.82cm},
  ttl/.style={font=\footnotesize\bfseries, align=center, inner sep=0pt, anchor=base},
  sub/.style={align=center, text width=2.55cm, inner sep=0pt, text=black!85},
  badge/.style={circle, fill=black!70, text=white, font=\scriptsize\bfseries,
    inner sep=0pt, minimum size=3.4mm},
  tip/.style={-{Stealth[length=1.6mm,width=1.3mm]}},
  seq/.style={tip, draw=black!40, line width=0.5pt},
  act/.style={tip, draw=osAccent, line width=0.8pt},
  alab/.style={font=\scriptsize\itshape, text=osAccent, inner sep=1.5pt},
]

\foreach \x in {0, 3.05, 6.1} {
  \draw[panel] (\x, 2.0) rectangle ++(2.7, 2.65);
}

\node[ttl] at (1.35, 4.19) {Capability rises};
\draw[black!45, line width=0.4pt] (0.45, 3.15) to (2.25, 3.15);
\foreach \i/\h in {0/0.22, 1/0.42, 2/0.62, 3/0.84} {
  \fill[black!35, rounded corners=0.6pt] (0.66 + 0.42*\i, 3.15) rectangle ++(0.26, \h);
}
\node[sub, anchor=north] at (1.35, 2.93) {so throughput $r$ rises};

\node[ttl] at (4.4, 4.19) {Knowledge decays};
\foreach \row in {0, 1} {
  \foreach \col in {0, ..., 4} {
    \fill[black!35, rounded corners=0.6pt] (3.41 + 0.42*\col, 3.17 + 0.42*\row) rectangle ++(0.3, 0.3);
  }
}
\foreach \col/\row in {1/0, 4/0, 3/1} {
  \filldraw[fill=white, draw=black!50, line width=0.45pt, dash pattern=on 1.2pt off 0.9pt,
    rounded corners=0.6pt] (3.41 + 0.42*\col, 3.17 + 0.42*\row) rectangle ++(0.3, 0.3);
}
\node[sub, anchor=north] at (4.4, 2.93) {some units go stale:\\probability $s$};

\node[ttl] at (7.45, 4.19) {Review is fixed};
\foreach \y in {3.38, 3.58, 3.78} {
  \draw[tip, draw=black!40, line width=0.5pt] (6.45, \y) to (7.38, \y);
}
\foreach \y in {3.15, 4.01} {
  \draw[tip, draw=black!85, line width=0.65pt] (6.45, \y) to (8.45, \y);
}
\fill[black!70, rounded corners=0.5pt] (7.42, 3.28) rectangle (7.52, 3.88);
\node[anchor=west, inner sep=1pt] at (7.56, 3.58) {$c$};
\node[sub, anchor=north] at (7.45, 2.93) {beyond $c$, stale-built\\changes escape};

\draw[seq] (2.73, 3.55) to (3.02, 3.55);
\draw[seq] (5.78, 3.55) to (6.07, 3.55);

\draw[gov] (0, 0) rectangle (8.8, 1.45);
\node[align=left, anchor=west, inner sep=0pt, text width=2.5cm] at (0.22, 0.7)
  {{\color{osAccent}\small\bfseries OSTEON}\\[1pt]governance built to\\scale with capability};
\node[mech] (dc) at (4.4, 0.7) {\textbf{Decay check}\\every session start};
\node[mech] (mg) at (7.45, 0.7) {\textbf{Mechanical gates}\\every change};
\draw[act] (dc.north) to node[alab, right, pos=0.6] {built to lower $s$} (4.4, 2.0);
\draw[act] (mg.north) to node[alab, right, pos=0.6] {$\rho=1$} (7.45, 2.0);

\foreach \x/\n in {0.3/1, 3.35/2, 6.4/3} {
  \node[badge] at (\x, 4.65) {\n};
}
\node[badge] at (0.3, 1.45) {4};
\end{tikzpicture}%%
}
\caption{Why capability is not enough (Section~\ref{sec:argument}). Capability raises throughput
$r$, a consulted knowledge unit is stale with probability $s$, and review capacity $c$ is fixed, so
once $r>c$ the share of stale-built changes that reading catches falls as $c/r$, and the escaping
number grows linearly with $r$ at the full rate of stale-built change (Eq.~\eqref{eq:escape}; a
model, not a measurement). OSTEON acts on the two factors governance acts on: its session-start
decay check is built to lower $s$, and its mechanical gates to check every change ($\rho=1$).}
\label{fig:teaser}
\end{figure}

Coding agents now resolve real issues in real repositories, and benchmarks such as
SWE-bench~\cite{jimenez2024swebench} track how often they succeed. One route to better software
engineering is to train more capable agents; we call it the \emph{training route}, and it now
states its goal plainly. Self-play SWE-RL trains a single agent to inject and repair bugs of rising
difficulty in real codebases, with no human-labelled issues or tests, and presents itself as a
first step toward training \emph{superintelligent software agents}~\cite{wei2026ssr}.
If that route succeeds, the agents that work in a codebase will, on much of the work, exceed the
engineers who are responsible for it. Fig.~\ref{fig:teaser} summarises the argument of this paper.

This paper asks what engineering discipline that regime requires. Our answer is that capability is
necessary and not sufficient. Software engineering has never been only the production of code that
passes its tests. It is the production of code that is correct with respect to what a project has
decided: the trade-offs it accepted, the contracts it holds with its neighbours, and the state of
the system it actually runs. Most of that knowledge lives outside the code. It changes, and parts
of it go stale without anyone noticing. In a snapshot of 452 of the most-starred public GitHub %%
repositories, 48.67\% carry a root-level or \texttt{.github} file written to brief coding agents. In the median such file that %%
names repository paths, between 9.5\% and 22\% of those paths do not resolve, depending on how a %%
path reference is read (Section~\ref{sec:measure}). An unresolved path is a mismatch between a
file and its tree, not shown to be a wrong instruction; it is one observable form of the drift this
paper is about. An agent that consults stale knowledge may build the wrong change, and cannot tell
that it has. We expect a more capable agent to build it faster, at larger scope, and to defend it
better. \op{}

Our thesis is therefore a scaling claim: \emph{governance has to scale with capability.} Four
things must grow in reach and fall in cost as agents grow in capability: the gates a change passes,
the contracts it is checked against, the decisions it must respect, and the control of how that
knowledge decays. If they do not, we argue, more capability produces more confident error.

\subsection*{Contributions}

\begin{enumerate}
\item A definition of superintelligent software engineering as a discipline, and of the regime it
addresses, distinguished from the agent that works in it and from the era that precedes it
(Section~\ref{sec:define}).
\item An argument that capability alone does not reduce the changes built on stale knowledge,
with a simple model of how such changes escape review as throughput rises
(Section~\ref{sec:argument}).
\item An exploratory measurement of agent context files in 452 of the most-starred public GitHub %%
repositories, released with its scripts and dataset so that its per-file statistics can be
recomputed (Section~\ref{sec:measure}).
\item Seven requirements that we propose for any discipline for the regime
(Section~\ref{sec:requirements}).
\item OSTEON, a concrete discipline built to these requirements, released as an open artifact with
a DOI~\cite{osteon2026zenodo} (Sections~\ref{sec:osteon} and~\ref{sec:scaling}).
\item A research agenda that states the thesis as testable claims, with the evidence that would
refute each, and one open design question (Section~\ref{sec:agenda}).
\end{enumerate}

\subsection*{Evidence status}

This is a vision paper with a solution and one exploratory measurement. It reports no measurement
of OSTEON and claims no measured effect. Section~\ref{sec:measure} examines one premise of the
argument, that written project knowledge drifts from the system it describes, in public agent
context files. It is a single observational snapshot: it describes, and it tests no cause.
Sections~\ref{sec:osteon} and~\ref{sec:scaling} describe mechanisms, and each can be checked
against the released artifact~\cite{osteon2026zenodo}. Where a sentence claims what a mechanism
achieves in practice rather than what it does, it is an opinion, and it is marked \op{} where it
stands. The scenarios in Section~\ref{sec:scaling} are constructed to show a mechanism; they are
not drawn from data. Section~\ref{sec:lessons} states design heuristics from the author's own
practice; each is marked \exper{} and is likewise not evidence.
Section~\ref{sec:osteon-history} gives a descriptive reading of OSTEON's own version history,
offered as a check on the mechanisms, not as a measurement of their effect.

\section{Defining the Discipline}
\label{sec:define}

\begin{quote}
\textbf{Definition.} \emph{Superintelligent software engineering} is the discipline of building and
evolving software with agents whose capability exceeds that of human engineers, in which governance
(gates, contracts, decisions and decay control) scales with capability.
\end{quote}

The definition is the one the OSTEON release carries~\cite{osteon2026zenodo}; this paper argues
it. Each of its three parts does work.

\emph{Building and evolving.} The definition covers the whole life of a system, not its first
release. Lehman observed that a program used in the real world must keep changing or become
progressively less useful~\cite{lehman1980laws}, and Parnas described how software ages as the
world around it moves~\cite{parnas1994aging}. The regime we address is mostly evolution: agents
changing systems whose decisions were taken earlier, by other people, for reasons that may no longer
hold.

\emph{Agents whose capability exceeds that of human engineers.} We do not require superintelligence
in the general sense of an intellect that exceeds human performance in virtually all domains of
interest~\cite{bostrom2014superintelligence}. We require something narrower and nearer: that on the
work delegated to it, an agent produces output that the project's engineers could not produce, or
could not verify line by line, in the time available. Capability is measured against the humans who
govern the project. The regime therefore arrives unevenly, task by task and project by project,
long before any general superintelligence.

The condition is relative, and it includes speed. The regime is defined by the gap between an
agent's output and the project's capacity to verify it, not by intelligence in the general sense: an
agent can enter it by depth, by scope or by speed. An agent no better than its reviewers but much
faster also produces more than they can verify in time. Capability enters the model of
Section~\ref{sec:arg-throughput} through throughput $r$ and scope $k$, where the two cases enter
alike, and through the detection probability $d$ where an agent's changes outrun a reader's depth;
the model holds $d$ fixed (Section~\ref{sec:limits}). We keep the term because the training route names
superintelligent agents as its target~\cite{wei2026ssr}, because depth is the case in which
verification fails even when there is time, and because we address the regime in which both the
quality and the volume of agent output exceed what reviewers can check; the fast but ordinary agent
is a transitional case that the same governance covers.

\emph{Governance scales with capability.} This is the defining property. A practice in which agents
become more capable while the governance around them stays fixed is not superintelligent software
engineering. It is ordinary software engineering run faster, and Section~\ref{sec:argument} argues
that it fails in a predictable way. A practice meets the definition when it meets the requirements
of Section~\ref{sec:requirements}; whether its governance then scales in fact is the empirical
question of Section~\ref{sec:agenda}.

The definition is stipulative: it names the discipline by the property we argue for, and does not
by itself show that the property is needed; a reader who rejects Section~\ref{sec:argument} can
accept the definition and hold the discipline empty. We keep apart the regime, a descriptive
condition on verification capacity (Section~\ref{sec:regimes}); the discipline, a normative
practice for it; and the claim that the practice is needed, which Section~\ref{sec:argument}
argues and Section~\ref{sec:agenda} says how to refute.

\subsection{Three regimes}
\label{sec:regimes}

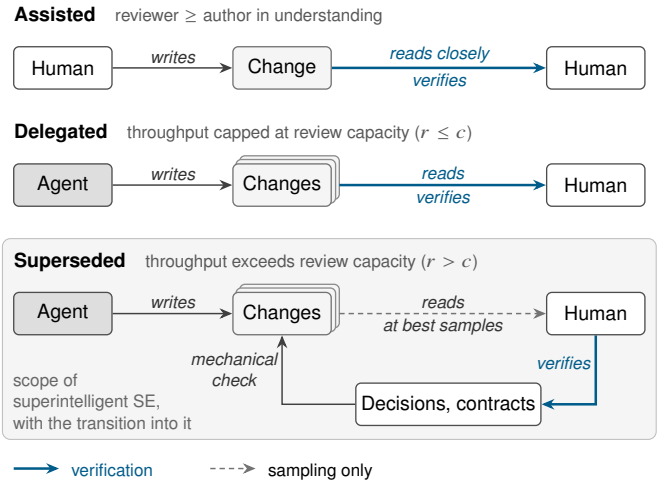
\begin{figure}[t]
\centering
{\sffamily\providecolor{osAccent}{RGB}{0,92,138}%%
\ifdefined\fontspec\else\fontencoding{T1}\fontfamily{phv}\selectfont\fi
\begin{tikzpicture}[
  font=\footnotesize,
  tip/.style={-{Stealth[length=1.7mm,width=1.4mm]}},
  actor/.style={draw=black!70, line width=0.5pt, rounded corners=2pt, fill=white,
    minimum width=1.3cm, minimum height=0.56cm, inner sep=2pt, align=center},
  human/.style={actor},
  agent/.style={actor, fill=black!13},
  chg/.style={actor, draw=black!55, fill=black!4},
  know/.style={actor, minimum width=2.3cm},
  flow/.style={tip, draw=black!75, line width=0.55pt},
  verify/.style={tip, draw=osAccent, line width=0.95pt},
  sample/.style={tip, draw=black!55, line width=0.55pt, dash pattern=on 2pt off 1.5pt},
  lab/.style={font=\scriptsize\itshape, text=black!80, inner sep=1.6pt, align=center},
  alab/.style={lab, text=osAccent},
  rhead/.style={anchor=base west, inner sep=0pt},
]

\def\osXa{0.65}\def\osXc{3.55}\def\osXh{7.7}
\def\osStack#1{%%
  \begin{scope}[on background layer]
    \foreach \o in {2.8pt,1.4pt} {
      \draw[draw=black!45, fill=black!4, line width=0.5pt, rounded corners=2pt]
        ([xshift=\o,yshift=\o]#1.south west) rectangle ([xshift=\o,yshift=\o]#1.north east);
    }
  \end{scope}}

\node[rhead] (na) at (0,0.62) {\textbf{Assisted}\quad
  {\scriptsize\color{black!65}reviewer $\geq$ author in understanding}};
\node[human] (aa) at (\osXa,0) {Human};
\node[chg] (ca) at (\osXc,0) {Change};
\node[human] (ha) at (\osXh,0) {Human};
\draw[flow] (aa) to node[lab, above] {writes} (ca);
\draw[verify] (ca) to node[alab, above] {reads closely} node[alab, below] {verifies} (ha);

\def\osYd{-1.55}
\node[rhead] (nd) at (0,\osYd+0.62) {\textbf{Delegated}\quad
  {\scriptsize\color{black!65}throughput capped at review capacity ($r\leq c$)}};
\node[agent] (ad) at (\osXa,\osYd) {Agent};
\node[chg] (cd) at (\osXc,\osYd) {Changes};
\node[human] (hd) at (\osXh,\osYd) {Human};
\osStack{cd}
\draw[flow] (ad) to node[lab, above] {writes} (cd);
\draw[verify] ([xshift=2.8pt]cd.east) to node[alab, above] {reads} node[alab, below] {verifies} (hd);

\def\osYs{-3.25}
\node[rhead] (ns) at (0,\osYs+0.62) {\textbf{Superseded}\quad
  {\scriptsize\color{black!65}throughput exceeds review capacity ($r>c$)}};
\node[agent] (as) at (\osXa,\osYs) {Agent};
\node[chg] (cs) at (\osXc,\osYs) {Changes};
\node[human] (hs) at (\osXh,\osYs) {Human};
\node[know] (ks) at (5.75,\osYs-1.2) {Decisions, contracts};
\draw[flow] (as) to node[lab, above] {writes} (cs);
\draw[sample] ([xshift=2.8pt]cs.east) to node[lab, above] {reads} node[lab, below] {at best samples} (hs);
\draw[verify] (hs.south) |- node[alab, left, pos=0.2] {verifies} (ks.east);
\draw[flow] (ks.west) -| node[lab, left, pos=0.75] {mechanical\\check} (cs.south);

\begin{scope}[on background layer]
  \node[fit=(ns)(as)(hs)(ks), inner xsep=4pt, inner ysep=5pt, rounded corners=3pt,
    fill=black!4, draw=black!25, line width=0.4pt] (scope) {};
\end{scope}
\osStack{cs}
\node[anchor=south west, font=\scriptsize, text=black!60, inner sep=4pt, align=left]
  at (scope.south west) {scope of\\superintelligent SE,\\with the transition into it};

\begin{scope}[shift={(0,-5.3)}, font=\scriptsize, every node/.append style={text height=1.5ex, text depth=0.3ex}]
  \draw[verify] (0,0) to (0.6,0); \node[right, text=osAccent] at (0.65,0) {verification};
  \draw[sample] (2.6,0) to (3.2,0); \node[right] at (3.25,0) {sampling only};
\end{scope}
\end{tikzpicture}%%
}
\caption{The three regimes, separated by who can verify a change. In the assisted and delegated
regimes a human verifies by reading the change (thick blue arrows mark where verification sits); in
the delegated regime review is the binding constraint. In the superseded regime the agent's output
exceeds what the humans responsible can read in the time they have, so reading at best samples
(dashed arrow); human verification moves to the decisions and contracts, and the step from that
knowledge to the change is mechanised (Section~\ref{sec:arg-oversight}). The shaded area is the
scope of superintelligent software engineering: the superseded regime and the transition into it.}
\label{fig:regimes}
\end{figure}

It helps to separate three regimes by who can verify a change (Fig.~\ref{fig:regimes}).

\begin{itemize}
\item \emph{Assisted.} A human writes or closely reads every change, and the reviewer understands
the change at least as well as its author (in the terms of Section~\ref{sec:arg-throughput}: every
change is reviewed and detection is high).
\item \emph{Delegated.} An agent writes and a human reads every change, which is possible only
because throughput is held at or below review capacity: review is the binding constraint ($\rho=1$
because $r$ is capped at $c$). Most current practice sits here. \op{} As we read them, the settings that
SE 3.0~\cite{hassan2024se3} and agentic software engineering~\cite{hassan2025agentic} describe sit
here too, because in both a human still reads and approves each change or its evidence pack.
\item \emph{Superseded.} Throughput exceeds review capacity, so reading samples ($\rho<1$), or the
reviewer can no longer judge a change even when reading it ($d$ falls).
\end{itemize}

Superintelligent software engineering is the discipline for the superseded regime and for the
transition into it. The transition is gradual and mixed: within one project, a schema migration may
be superseded while a pricing rule is still assisted. A project can stay delegated by capping its
agents at review capacity. That forgoes the capability. The superseded regime is entered by projects
that do not cap it, and our claim is conditional on entering it.

\subsection{Why name a discipline}

Recent work has named the agent (Self-play SWE-RL~\cite{wei2026ssr}), the era (SE 3.0, an
AI-native era with its technology stack~\cite{hassan2024se3}), the human-agent duality (agentic
software engineering, with workbenches for SE for Humans and for SE for
Agents~\cite{hassan2025agentic}) and the trust problem (Roychoudhury et al., who argue that
agent-written code needs trust at least equal to that placed in human practice and look to analysis
tools paired with the agent to supply it~\cite{roychoudhury2026trust}), but not the discipline. We
use \emph{discipline} in the sense of a defined body of practice, not of an established research
community. None of these works makes the scaling of governance the defining property of a practice
for the regime in which agents outstrip their reviewers. We are not aware of an earlier definition
of superintelligent software engineering as a discipline; a search for the phrase on 2026-10-17
found it only as the target of agent training~\cite{wei2026ssr}.

\section{Why Capability Is Not Enough}
\label{sec:argument}

The training route improves what an agent can do with the knowledge it has. Three claims explain why that is not the same as engineering a project well: correctness,
decay and throughput (Sections~\ref{sec:arg-correct} to~\ref{sec:arg-throughput}). Two further
points say what follows for oversight and for authority (Sections~\ref{sec:arg-oversight}
and~\ref{sec:arg-authority}).

\subsection{Correctness is relative to knowledge the code does not carry}
\label{sec:arg-correct}

A change is correct relative to an intent. Some intent is written in tests, and the training route
uses exactly that: in self-play SWE-RL each bug is specified by a test
patch~\cite{wei2026ssr}. Much of a project's intent, however, is neither in its tests nor
recoverable from its code: why a retry is capped where it is; which team owns a field, and in what
unit; which approach was tried and rejected; which part of the system is in production and which
is a prototype. These are decisions and contracts. Architecture research has long treated decisions as first-class knowledge that the code does not
preserve~\cite{jansen2005decisions,tyree2005decisions}. The lightweight decision record, a short
written record of one decision and its reasons, exists because that rationale is otherwise
lost~\cite{nygard2011adr}. A more capable agent infers more from code. It cannot infer a decision
that left no trace, and it cannot know that a decision it does infer has since been reversed.

\subsection{Project knowledge decays}
\label{sec:arg-decay}

The knowledge that does get written down drifts from the system it describes. Documentation that
was true when written goes stale as the code, the deployment and the organisation move; code and
its comments co-evolve only loosely~\cite{fluri2007coevolve}, code-comment inconsistencies are
common in mature projects~\cite{wen2019inconsistencies}, and outdated content is among the most
reported documentation problems~\cite{aghajani2019docissues}. A stale fact can be worse than a
missing one: an agent treats it as current and acts on it with full confidence. \op{} Section~\ref{sec:measure} measures one possible symptom of this drift in public
repositories.

A recent controlled evaluation of repository context files bears on this. Gloaguen et al.\ found
that providing context files such as \texttt{AGENTS.md} does not generally improve task success,
while it raises inference cost by about a fifth. Instructions in the files are followed,
but repository overviews do not help~\cite{gloaguen2026agentsmd}. Gloaguen et al.\ themselves
conclude that such files are useful for specifying non-standard coding practices. Our own addition,
not theirs, is that the lever is not more context but current context: what an agent cannot derive
from the code is a small set of non-standard, project-specific facts, and their value depends on
whether they are current. \op{} Context that is ungoverned, unversioned and unchecked for decay is cost
with no assurance of benefit.

\subsection{Capability amplifies the rate of stale-built change}
\label{sec:arg-throughput}

A simple model makes the scaling explicit. It is a model, not a measurement.

\begin{figure}[t]
\centering
\includegraphics[width=\columnwidth]{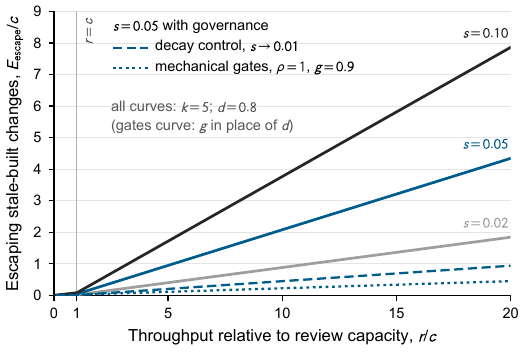}
\caption{The escape model of Eq.~\eqref{eq:escape}, normalised by review capacity:
$E_{\text{escape}}/c$ against $r/c$, with $\rho=\min(1,c/r)$, $k=5$ and $d=0.8$. Solid lines show
three staleness levels $s$ under review by reading alone. Below $r=c$ (grey vertical line) every
change is reviewed and the escape rate grows at slope $(1-d)$ times the stale fraction; above it
the reviewed fraction falls as $c/r$ and the slope rises to the full stale fraction. Broken lines show the $s=0.05$ case with governance: decay control lowers
$s$ to $0.01$ (dashed), and mechanical gates check every change ($\rho=1$) with detection
probability $g=0.9$ in place of $d$ (dotted). All parameter values are illustrative, not measured;
the script and parameters ship with the figure.}
\label{fig:escape}
\end{figure}

Let an agent make changes at rate $r$. Each change consults $k$ units of project knowledge (a
decision, a contract clause, a state record). Let $s$ be the probability that a consulted unit is
stale, that is, that its source has changed since it was last verified. If staleness is
independent across units, the rate of changes that rest on at least one stale unit is
\begin{equation}
E_{\text{stale}} = r\,\bigl[1-(1-s)^{k}\bigr].
\label{eq:stale}
\end{equation}
A stale unit need not be wrong, so $E_{\text{stale}}$ bounds, rather than counts, the changes that
knowledge decay makes wrong. Suppose humans can review changes at a fixed capacity $c$, so the fraction reviewed is
$\rho=\min(1,c/r)$, and a review catches a stale-built change with probability $d$. The rate of
stale-built changes that escape review is then
\begin{equation}
E_{\text{escape}} = r\,\bigl[1-(1-s)^{k}\bigr]\,\bigl(1-\rho\,d\bigr).
\label{eq:escape}
\end{equation}
A stale-built change is one exposed to a stale unit, and only some fraction of them are wrong; a
constant factor for that fraction would scale every curve in Fig.~\ref{fig:escape} and change none
of the conclusions, so we omit it. Fig.~\ref{fig:escape} plots Eq.~\eqref{eq:escape} for
illustrative parameter values.

Three consequences follow. First, capability raises $r$ and, we assume, $k$, because more
ambitious changes consult more knowledge; both raise $E_{\text{escape}}$. Second, the escape rate
grows linearly in throughput on both sides of $c$. Below $c$ its slope is $(1-d)$ times the stale
fraction $1-(1-s)^{k}$, because a reader misses some stale-built changes even when every change is
read. Above $c$ the reviewed fraction $\rho$ falls as $c/r$ and
$E_{\text{escape}} = [1-(1-s)^{k}]\,(r - cd)$, whose slope is the whole stale fraction, as if no
review existed. The threshold at $r=c$ changes the slope, not the kind of growth; it is where
reading stops mattering, and for any fixed $s>0$ the growth is without bound. Third, the model holds
$s$ outside the agent: staleness is a property of the knowledge, not of the agent, so training does
not act on it. For a claim that the code does not carry (a decision, a contract with another team,
the state of production), an agent can detect staleness only if the unit carries provenance (a
record of the sources it came from)~\cite{moreau2013prov} and changes to those sources can be
observed. That metadata is not in the code.

Of the terms in Eq.~\eqref{eq:escape}, capability acts on $r$ and $k$. Governance acts on two
factors without adding reviewers: the staleness factor, through $s$, and the review factor, through
$\rho$ and $d$. Decay control lowers $s$ by finding stale units before they are consulted. Gates act
on the review factor. By \emph{mechanical} we mean run on every change at no human cost. A gate runs
on every change, so it sets $\rho$ to 1, and it holds a change that rests on a unit already flagged
as stale with its own probability $g$ in place of a reader's $d$ (Fig.~\ref{fig:escape}). Its other
job, catching a change that contradicts a current record, is a second error channel that
Eq.~\eqref{eq:escape} does not count (the second scenario of Section~\ref{sec:scaling} shows it).
The two factors are not independent: a gate cannot detect a record that is itself stale and
unflagged, and a stale-built change that agrees with such a record passes it. The gate's $g$ is therefore
bounded by the currency of the record it checks against, which is why decay control comes first,
and why the gate curve in Fig.~\ref{fig:escape} assumes a current record. In OSTEON 0.1.0 the decay
check and the lints are scripts; the review gate is a procedure the agent executes against the
record, so its $g$ is a property of the agent and the record together, which is why Q5 of
Section~\ref{sec:agenda} asks whether it holds as capability rises.

A project could also hold $r$ at $c$, which is the delegated regime's answer and forgoes the
capability, or raise $c$: with more human reviewers, which is what the superseded regime by
definition can no longer do, or with agent reviewers, which the fifth objection of
Section~\ref{sec:objections} takes up (an agent that reviews against the record is, in our terms,
a gate). Both factors are governance, and neither improves by itself when the agent improves. This
is the sense in which governance has to scale with capability.

\subsection{Verification moves from the change to the knowledge}
\label{sec:arg-oversight}

When the supervised system exceeds its supervisor, oversight has to be designed so that the
supervisor can still judge something that matters. This is the problem of scalable
oversight~\cite{amodei2016concrete,bowman2022scalable}. In the superseded regime the reviewer
cannot judge every change. The reviewer can still judge the decisions and contracts the change is
checked against, because those are short, stable, human-approved and far fewer than the changes. \op{}
Superintelligent software
engineering moves human verification from the change to the knowledge, and mechanises the step
from knowledge to change.

\subsection{Authority does not follow capability}
\label{sec:arg-authority}

A decision records a trade-off among things a project's owners answer for: cost, risk, obligations
to users and to other teams. An agent may predict the consequences of a decision better than any
engineer and still not be the party entitled to take it. We hold that the boundary of human
authority should not move as capability grows; what should fall is the cost of exercising it. \op{}

\subsection{The scaling principle}
\label{sec:arg-scaling}

Together these give the principle the definition names. We say governance scales with capability
when five conditions hold as capability rises. Two are scaling properties in the strict sense, cost
and coverage; the other three, currency, authority and learning, are invariants that must survive
the scaling. In the model's terms, governance scales when $\rho$ stays at 1 and $s$ does not rise as
$r$ rises.

\begin{itemize}
\item \emph{Cost.} Governance cost per change falls as change rate rises, so it is not skipped
under load. Checks that can be mechanical are mechanical.
\item \emph{Coverage.} What is governed grows with the scope agents take on. Every durable choice
an agent relies on is recorded, and every seam it crosses has a contract.
\item \emph{Currency.} Every governed fact carries its provenance. Decay is checked before work
starts rather than discovered after a failure.
\item \emph{Authority.} Agents propose; accountable humans dispose: they make the final call on anything
that changes a decision or a contract. Human effort attaches to decisions, which are rare, not to changes, which
are many.
\item \emph{Learning.} The process improves from what sessions learn, without any session
legislating for the next.
\end{itemize}

\section{Context Decay in Public Repositories: An Exploratory Measurement}
\label{sec:measure}

Section~\ref{sec:argument} rests on the premise that written project knowledge drifts from the
system it describes. Agent context files are a common way to give coding agents written project
knowledge, and in public repositories they can be observed directly, so we took one snapshot of
them. The measurement concerns the premise, not OSTEON. It is exploratory and
observational: it tests no hypothesis and supports no causal claim. It asks three measurement
questions (MQ1 to MQ3).

\begin{description}[\IEEEsetlabelwidth{MQ3}]
\item[MQ1] \emph{Prevalence.} How many popular repositories carry a root-level or \texttt{.github}
agent context file?
\item[MQ2] \emph{Currency.} How long ago was each file last updated, and how many commits has its
repository received since?
\item[MQ3] \emph{Resolution.} What share of the repository paths a file names do not resolve in the
repository tree at the snapshot?
\end{description}

Earlier empirical studies characterise the content, adoption and evolution of context files in
GitHub repositories~\cite{chatlatanagulchai2025manifests,mohsenimofidi2025context,chatlatanagulchai2025readmes};
MQ3 applies a link check, routine in documentation pipelines, to every path a context file names,
against its repository's tree.

\subsection{Method}
\label{sec:measure-method}

\emph{Frame.} We queried the GitHub search API for repositories matching
\par\smallskip\centerline{\texttt{stars:>10000 fork:false archived:false}}\smallskip\noindent %%
sorted by stars in descending order, and took the first five pages of 100 results. The query %%
matched 5\,346 repositories, and the five pages returned %%
500, none marked incomplete and none duplicated across pages (Table~\ref{tab:funnel}). We excluded %%
47 repositories with no primary language, a rule meant to remove lists and books. One repository %%
matched a list, fixed before collection, of names the author does not report on for reasons of
confidentiality; it was excluded, and it is counted but not named. All 452 %%
eligible repositories were measured, no repository tree was unavailable, and every context %%
file's content and path history were available. For 7 repositories %%
the recursive tree listing was truncated, so the context-file paths were checked one by one
through the contents API. The snapshot date, \mbox{2026-10-18} (UTC), is the reference for every %%
age; collection used read-only endpoints only. The eligible repositories hold between 50\,069 and %%
554\,124 stars and are assigned 32 primary languages; Python (130), TypeScript (85), JavaScript %%
(50), Rust (33) and Go (32) account for 330 of the 452. The frame is released with the dataset %%
(\texttt{frame.csv}).

\emph{Context files.} At each repository's default branch we looked for five file names, matched
exactly and case-sensitively (Table~\ref{tab:funnel}):
\begin{filelist}
\item at the repository root: \texttt{AGENTS.md}, \texttt{CLAUDE.md}, \texttt{GEMINI.md},
\texttt{.cursorrules};
\item in \texttt{.github}: \texttt{copilot-instructions.md}.
\end{filelist}
\texttt{AGENTS.md} or \texttt{CLAUDE.md} files elsewhere in the tree are \emph{nested} files; we
counted them but did not analyse them. The \emph{sample} is the set of repositories with at least
one root-level or \texttt{.github} context file, and each such file is one unit of analysis.

\emph{Pre-specification.} The frame, the file list and the metrics were written down verbatim, in
the study's protocol and in the header of the collection script, before collection began. One
robustness change was made after a small test on a separate cache and before the full run: the tree
call was set to tolerate two further HTTP error responses (403 and 451); the test data were not
used, the protocol records the change in a dated note, and no tree call in the full run failed, so %%
the change affected no result. The summary statistics (medians and quartiles, the pooled rate, the
share of files with any unresolved reference) and the single association reported, commits since
update against the dangling-reference rate, were implemented in \texttt{summarize.py} before
collection and were not changed afterwards; every other comparison in this section is labelled
post hoc, including the audit in Section~\ref{sec:measure-sens} and the intervals and breakdowns
added at revision. The pre-specification is recorded only in the released files; it was not
registered with a third party, so it rests on the author's word, the file record and the dated
deviation note.

\begin{table}[t]
\caption{Collection funnel. Every count is read from the study's funnel record; counts that are
zero (duplicates across pages, trees unavailable, files unreadable or without path history) are
stated in the text.}
\label{tab:funnel}
\centering
\footnotesize
\begin{tabularx}{\columnwidth}{@{}Y r@{}}
\toprule
Stage & Count \\
\midrule
\multicolumn{2}{@{}l}{\emph{Repositories}} \\
Matching the search query & 5\,346 \\ %%
Returned by search (five pages of 100) & 500 \\ %%
\quad Excluded: no primary language & 47 \\ %%
\quad Excluded: privacy denylist (counted, not named) & 1 \\ %%
Eligible and measured & 452 \\ %%
\quad Truncated tree, checked through the contents API & 7 \\ %%
Sample: a root-level or \texttt{.github} context file & 220 \\ %%
Nested files only (outside the sample) & 14 \\ %%
\midrule
\multicolumn{2}{@{}l}{\emph{Context files in the sample}} \\
All root-level and \texttt{.github} files & 367 \\ %%
\quad \texttt{AGENTS.md} & 189 \\ %%
\quad \texttt{CLAUDE.md} & 112 \\ %%
\quad \texttt{.github/copilot-instructions.md} & 49 \\ %%
\quad \texttt{GEMINI.md} & 15 \\ %%
\quad \texttt{.cursorrules} & 2 \\ %%
Stored as symbolic links & 35 \\ %%
With at least one path-like candidate & 269 \\ %%
\bottomrule
\end{tabularx}
\end{table}

\subsection{Metrics}
\label{sec:measure-metrics}

For each context file in the sample we compute five measures.

\begin{itemize}
\item \emph{Age.} Days from the oldest commit that touches the file to the snapshot date.
\item \emph{Days since update.} Days from the newest commit that touches the file to the snapshot date.
\item \emph{Updates.} The number of default-branch commits that touch the file's path. GitHub's path
history does not follow renames, so a moved file is aged from the move.
\item \emph{Commits since update.} The number of default-branch commits dated at or after the file's last
update, excluding that update. It measures how far the repository moved while the file stood still.
\item \emph{Dangling-reference rate.} The share of a file's distinct path-like references that do not
resolve in the tree at the snapshot. Candidates are inline code spans and Markdown link targets that look like repository paths. A
candidate contains ``/'' or ends in an extension from a fixed list. It contains no whitespace, is
not a URL, and passes the further filters of the released protocol (\texttt{README.md}, operational
detail 6). Fenced code blocks are removed first. A candidate resolves if it
equals a path in the tree, or is a directory prefix of one, either from the repository root or
relative to the context file's directory. The rate is undefined for a file with no candidates.
\end{itemize}

Prevalence is the number of sample repositories over the number of eligible repositories. Every
distribution is skewed, so we report medians with 25th and 75th percentiles (linear interpolation)
and use the Spearman rank correlation, $r_{\text{S}}$, for association. Where a confidence interval
is given it is a Wilson 95\% interval for a proportion or a seeded percentile bootstrap (10\,000 %%
resamples) for a median; the study is descriptive, so we report intervals and no tests. An
unresolved path is a mismatch between a file and its current tree. It may reflect drift, a
generated or planned path, or the extraction heuristic. It is not proof that an instruction is
wrong.

\subsection{Results}
\label{sec:measure-results}

\emph{MQ1: Prevalence.} Of the 452 eligible repositories, 220 (48.67\%; 95\% confidence interval %%
44.1 to 53.3; 44.0\% over all 500 returned repositories) carry at least one %%
root-level or \texttt{.github} context file. Prevalence varies by language: TypeScript 62 of 85 and %%
Go 23 of 32, against Python 53 of 130, JavaScript 19 of 50 and Java 4 of 15 (post hoc breakdown). %%
Together the sample repositories hold 367 files, a median of 2 per %%
repository (25th percentile 1, 75th percentile 2, maximum 4). The most frequent names are %%
\texttt{AGENTS.md} (189 files) and \texttt{CLAUDE.md} (112 files); Table~\ref{tab:funnel} gives %%
the full breakdown. Of the 367 files, 35 are stored as symbolic links. Nested \texttt{AGENTS.md} %%
or \texttt{CLAUDE.md} files are common as well: 102 repositories hold 1\,299 of them, 88 of these %%
repositories (1\,277 files) in the sample and 14 outside it. %%

\begin{figure}[t]
\centering
\includegraphics[width=\columnwidth]{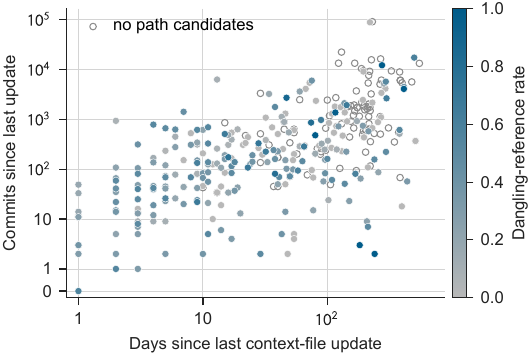}
\caption{Currency of the 367 context files: commits to the repository since the file's last %%
update, against days since that update. Both axes are logarithmic; the vertical axis is linear
below 1 because zero counts occur. Filled markers are files with at least one path-like candidate,
shaded by dangling-reference rate; hollow markers are files with none. The plot is descriptive.}
\label{fig:drift}
\end{figure}

\emph{MQ2: Currency.} Table~\ref{tab:stats} and Fig.~\ref{fig:drift} summarise the per-file measures. Context files have a
median age of 220 days (25th to 75th percentile 144 to 324.5) and a median of 48 days since their last %%
update (10 to 167). They have a median of 5 updates (1 to 18). Since their last update, their %%
repositories have received a median of 229 further commits (49 to 852). The %%
distribution has a heavy upper tail, reaching 91\,084 commits, because the count includes every %%
commit dated after the update, merged branches included (bootstrap 95\% interval for the median, %%
168 to 307). The 35 symbolic links are rows whose text is a target name, so they need no update when %%
their target changes; over the 332 other files the medians are 40 days since update (25th to 75th %%
percentile 9 to 134) and 197.5 commits since update (45 to 663.5), and the age median is 219.5, %%
against 220 over all files. By file name (post hoc), \texttt{AGENTS.md} files are the most current (median 17 days %%
since update, 125 commits since) and \texttt{CLAUDE.md} files the least (124.5 days, 457 commits); %%
median dangling rates are 0.22 and 0.27, and 0.15 for \texttt{copilot-instructions.md}. The cell %%
counts for \texttt{GEMINI.md} (15 files, 3 with candidates) and \texttt{.cursorrules} (2 files) %%
are too small to read.

\begin{table}[t]
\caption{Per-file summary statistics (pre-specified protocol). Age and days since update are in
days at the snapshot date; files, updates and commits are counts; the dangling-reference rate is a
fraction of a file's candidates.}
\label{tab:stats}
\centering
\footnotesize
\setlength{\tabcolsep}{3.6pt}
\begin{tabularx}{\columnwidth}{@{}Y rrrrrr@{}}
\toprule
Measure & $n$ & Median & P25 & P75 & Min & Max \\
\midrule
Files per sample repository & 220 & 2 & 1 & 2 & 1 & 4 \\ %%
Age & 367 & 220 & 144 & 324.5 & 12 & 689 \\ %%
Days since update & 367 & 48 & 10 & 167 & 1 & 549 \\ %%
Updates & 367 & 5 & 1 & 18 & 1 & 631 \\ %%
Commits since update & 367 & 229 & 49 & 852 & 0 & 91\,084 \\ %%
Dangling-reference rate & 269 & 0.2222 & 0 & 0.4231 & 0 & 1 \\ %%
\bottomrule
\end{tabularx}
\end{table}

\begin{figure}[t]
\centering
\includegraphics[width=\columnwidth]{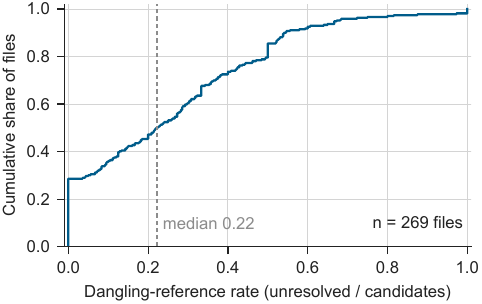}
\caption{Empirical cumulative distribution of the dangling-reference rate over the 269 context %%
files with at least one path-like candidate (pre-specified reading). The dashed line marks the
median. The step at zero is the share of files whose candidates all resolve.}
\label{fig:ecdf}
\end{figure}

\emph{MQ3: Resolution.} Of the 367 files, 269 contain at least one path-like candidate (Fig.~\ref{fig:ecdf}); none of %%
the 35 symbolic links has one, because a link's text is its target. Among these 269 files, the %%
median dangling-reference rate is 0.22 (25th to 75th percentile 0 to 0.42; bootstrap 95\% %%
interval 0.18 to 0.27). Of these files, 192 (71.38\%; 95\% confidence %%
interval 65.7 to 76.4) have at least one unresolved reference. Pooled over files, 2\,003 of %%
6\,172 candidates (32.45\%) do not resolve. Commits since update and the dangling-reference rate %%
are not positively associated ($r_{\text{S}}=-0.15$, $n=269$ files). Files cluster within %%
repositories (112 of the 220 sample repositories contribute more than one file), and at repository %%
level, with candidates pooled within each of the 205 repositories that have any, the median rate is %%
0.27 and the coefficient against the largest commits-since-update among a repository's files is %%
0.03 (post hoc). Files that have stood still through more commits do not show more unresolved %%
references in this sample at either level. We read nothing causal into this: repositories differ in
size, activity and the way their files refer to paths.

\subsection{Sensitivity of the dangling-reference rate (post hoc)}
\label{sec:measure-sens}

After seeing the primary results, we audited the 2\,003 unresolved references and, after %%
inspecting the most frequent unresolved tokens, defined two classes that are unlikely to be meant
as tracked repository paths: bare extensions such as \texttt{.ts} (97), and %%
paths under build or local folders such as \texttt{dist} or \texttt{node\_modules}, which
repositories usually do not commit (95). Together they make up 9.59\% of unresolved references. %%
Of the remaining 1\,811, 1\,115 (61.57\%) are bare file names with no ``/''. The protocol resolves %%
such names only at the root or next to the context file, so a file of that name deeper in the
tree counts as unresolved.

Table~\ref{tab:sensitivity} reports the measures under two post hoc readings. Reading A drops the
two classes. Reading B also keeps only candidates that contain ``/''. The median rate falls from
0.2222 to 0.1818 under A and to 0.0952 under B. The share of files with any unresolved reference falls %%
from 71.38\% to 68.28\% and 61.44\%. The headline rate therefore depends strongly on how %%
references are read.

\begin{table}[t]
\caption{Dangling-reference measures under the pre-specified reading and two post hoc readings.
A: without bare extensions and conventionally untracked build or local paths. B: as A, and only
candidates that contain ``/''. B also leaves out the three withheld candidates, whose form is
unknown.}
\label{tab:sensitivity}
\centering
\footnotesize
\setlength{\tabcolsep}{4pt}
\begin{tabularx}{\columnwidth}{@{}Y rrr@{}}
\toprule
Measure & Pre-specified & A (post hoc) & B (post hoc) \\
\midrule
Files with candidates & 269 & 268 & 236 \\ %%
Median rate & 0.2222 & 0.1818 & 0.0952 \\ %%
25th percentile & 0 & 0 & 0 \\ %%
75th percentile & 0.4231 & 0.3917 & 0.2857 \\ %%
Files with any unresolved & 192 & 183 & 145 \\ %%
\quad as \% of files & 71.38 & 68.28 & 61.44 \\ %%
Candidates, pooled & 6\,172 & 5\,940 & 3\,705 \\ %%
\quad unresolved & 2\,003 & 1\,811 & 696 \\ %%
\quad as \% of candidates & 32.45 & 30.49 & 18.79 \\ %%
\bottomrule
\end{tabularx}
\end{table}

\subsection{What the measurement shows, and what it does not}
\label{sec:measure-discussion}

The snapshot shows that the mismatch the premise of Section~\ref{sec:argument} predicts is common;
it does not show how it arose. About half of the popular repositories we measured brief their
agents through context files. For half of those files, the repository has received 229 or more
commits since the file last changed. Under all three %%
readings, most files that name repository paths name at least one that does not resolve. The one
association we examined does not support a simple accumulation story: files that stood still %%
through more commits do not have more unresolved references, at file or at repository level.

The measurement does not show that unresolved references mislead agents, that they cause defects,
or that any governance practice, OSTEON included, would reduce them. Nor is it a judgement of any
project: a reference that does not resolve may name a file the build generates, a path in another
repository, or a file that is planned, and whether a given referent moved, was never created, or is
generated is not determined. Its role is narrower: written path references that do not resolve
against the current tree are common in public repositories today, and they are one observable form
of the mismatch that R5 (decay is checked at the start) targets. It supplies a cross-sectional
reference point for Q3 of the research agenda (Section~\ref{sec:agenda}): a level at one date, not
a decay rate.

\subsection{Threats to validity}
\label{sec:measure-threats}

\emph{Construct validity.} An unresolved reference is a symptom, not a defect. Candidates are found
by pattern, not by understanding. Some flagged strings are not meant as paths, such as a package
name that ends in a file extension, a path the build creates, a Git reference such as
\texttt{origin/main}, a branch-name prefix such as \texttt{feat/} read as a directory, or the
literal \texttt{N/A}; the released candidate list shows each one, and the post hoc readings remove
none of these when the token contains a slash. Some real references are missed, such as paths in
prose without code formatting, paths inside fenced blocks, and bare names without an extension. The
post hoc readings bound part of this effect, not all of it. The rate rises with the number of
candidates a file holds ($r_{\text{S}}=0.41$ over the 269 files; candidates per file have a median %%
of 13, 25th to 75th percentile 4 to 31; of the 35 files with a single candidate, 3 have it %%
unresolved). This is consistent with more drift in path-heavy files and with a heuristic whose
false-positive share grows with file length, and the snapshot cannot separate the two. Commits since update
counts every commit dated after the update, merged branches included, so it can overstate activity;
we report medians for this reason. For 2 files the context-file commit was not found on the last %%
page of the activity range, so their count is not reduced by one.

\emph{Internal validity.} The frame, the file list, the metrics and the summary statistics were
fixed before collection; the sensitivity audit and the breakdowns added at revision were not, and
they are labelled post hoc. The 35 symbolic links, whose text is %%
the link target, enter the age and activity statistics, where they are the stalest rows;
Section~\ref{sec:measure-results} gives the medians without them. They have no candidates and so no
dangling-reference rate. The unit of analysis is the file, and files are not independent across a
repository; the repository-level figures in Section~\ref{sec:measure-results} are the check. For
the 7 repositories with truncated %%
trees, the nested-file counts are lower bounds.

\emph{External validity.} The frame is the most-starred non-fork, non-archived repositories, all
above 50\,000 stars; neither the prevalence nor the resolution figures generalise to GitHub as a %%
whole, still less to private projects. The no-primary-language rule removes only repositories to
which GitHub assigns no language; a few documentation repositories with a Markdown primary language
remain eligible (4 of 452, none in the sample), which can only lower the measured prevalence. File %%
names are matched exactly; lower-case variants and per-directory rule folders are outside the
protocol. Nested files are counted but not analysed.

\emph{Reliability.} Search ranking and repository contents change, so a rerun on a later date will
differ. The scripts, the per-file data and the per-candidate evidence are released, so every
per-file statistic can be recomputed; the funnel counts are in \texttt{funnel.json}, and rerunning
\texttt{collect.py} repeats the collection against the live API. Three candidate strings that matched the confidentiality list are %%
withheld in the released candidate list; they keep their resolution flag and enter the
pre-specified reading and reading A, not reading B.

\section{Requirements for a Discipline}
\label{sec:requirements}

The four objects the thesis names (gates, contracts, decisions and decay control) and the five
conditions of Section~\ref{sec:arg-scaling} become testable when stated as requirements on a
practice. We propose seven. They were written together with OSTEON, and Table~\ref{tab:mapping}
shows the correspondence; we do not claim they were derived independently of the design, and a
reader should test them against other practices.

\begin{description}[\IEEEsetlabelwidth{R7}]
\item[R1] \emph{One owner per fact.} Each fact lives at exactly one level of an explicit hierarchy, the
narrowest that fully owns it, and other levels reference it rather than copy it. Copies are where
decay hides. \op{}
\item[R2] \emph{Decisions are records.} Durable choices are versioned records with a lifecycle (proposed,
accepted, superseded), amended or superseded and never silently rewritten.
\item[R3] \emph{Seams have contracts.} Every interface between modules has a versioned contract that
changes together with the interface, never after it.
\item[R4] \emph{Gates take their bar from the project.} Review checks a change against the project's own
decisions, contracts and rules, not against generic best practice, and detects when a change
invalidates one of them.
\item[R5] \emph{Decay is checked at the start.} Every governed claim carries provenance to its sources,
and a check at the start of every session reports which claims rest on sources that have changed.
\item[R6] \emph{Change is planned and gated.} Work of any size is planned, the plan passes an explicit
gate before execution, and a definition of done states what proves completion.
\item[R7] \emph{Learning without self-legislation.} Sessions record what they learn, but no session
writes the rules that govern the next; a rule changes only through corroboration and human
approval.
\end{description}

R1 to R3 address coverage. R4 and R5 address currency and the detection terms of
Eq.~\eqref{eq:escape}. R6 addresses authority and cost, and R7 addresses learning. A practice that
meets all seven still needs evidence that it helps; Section~\ref{sec:agenda} says what that
evidence would look like.

\section{OSTEON}
\label{sec:osteon}

OSTEON is a proposed discipline for superintelligent software engineering. Release 0.1.0 is a
plugin for the Claude Code agent. It contains fifteen skills (procedures the agent follows; the
source repositories hold further skills that the release does not ship), a scaffold for the project
knowledge repository, and a session-start decay check. Content is under CC BY 4.0 and scripts under
MIT~\cite{osteon2026zenodo}. Fig.~\ref{fig:architecture} shows the parts described in this section
and how they connect.

\begin{figure*}[t]
\centering
{\sffamily\definecolor{osAccent}{RGB}{0,92,138}%%
\begin{tikzpicture}[
  font=\footnotesize,
  box/.style={draw=black!70, line width=0.5pt, rounded corners=1.5pt, fill=white,
    align=left, text width=#1, inner sep=3pt, minimum height=0.95cm},
  box/.default=3.1cm,
  lay/.style={draw=black!40, line width=0.4pt, fill=black!7, align=left,
    text width=5.75cm, inner sep=2.4pt, minimum height=0.48cm},
  tip/.style={-{Stealth[length=1.9mm,width=1.5mm]}},
  flow/.style={tip, draw=black!75, line width=0.6pt},
  check/.style={tip, draw=black!75, line width=0.6pt, dash pattern=on 2.4pt off 1.6pt},
  approve/.style={tip, draw=osAccent, line width=0.95pt},
  lab/.style={font=\scriptsize\itshape, text=black!85, inner sep=1.6pt, align=center},
  alab/.style={lab, text=osAccent},
]

\node[box=5.85cm] (t0) at (8.7,0)
  {\textbf{T0 organisation tier}\\methodology; register of named methodologies};

\node[lay] (l1) at (8.7,-2.1) {\textbf{L1}\enspace constitution};
\node[lay, below=2pt of l1] (l2) {\textbf{L2}\enspace decision records; dated state records};
\node[lay, below=2pt of l2] (l3) {\textbf{L3}\enspace versioned contracts; domain knowledge};
\node[lay, below=2pt of l3] (l4) {\textbf{L4}\enspace project skills (procedures)};
\node[lay, below=2pt of l4, fill=white, dash pattern=on 1.6pt off 1.2pt] (man)
  {\textbf{Manifest}\enspace declares which files are knowledge units and which sources are watched};
\node[above=2pt of l1, inner sep=1pt, align=center] (bonehdr)
  {\textbf{T1 project tier: the \textsc{bone}}\\a repository of knowledge; no product code};

\node[box=5.85cm] (t2) at (8.7,-6.25)
  {\textbf{T2 repository tier}\\code repositories; conventions in each one's agent instruction file};

\begin{scope}[on background layer]
  \node[fit=(t0)(bonehdr)(man)(t2), inner xsep=5pt, inner ysep=5pt, rounded corners=3pt,
    fill=black!3, draw=black!25, line width=0.4pt] (tiers) {};
  \node[fit=(bonehdr)(l1)(man), inner sep=3.5pt, rounded corners=2pt,
    fill=white, draw=black!70, line width=0.7pt] (bone) {};
\end{scope}
\node[anchor=south, font=\scriptsize, text=black!60, inner sep=1.5pt] at (tiers.north)
  {one owner per fact: the narrowest tier that fully owns it};

\node[box] (cap) at (1.65,0)
  {\textbf{Capture}\\a ledger that governs nothing; nominates on corroboration};
\node[box, draw=black!85, line width=0.8pt] (agent) at (1.65,-1.8)
  {\textbf{Agent session}\\reads, compares, drafts and proposes};
\node[box] (oss) at (1.65,-3.55)
  {\textbf{Ossification}\\planned, gated change: daily cycle $\rightarrow$ start gate (ten conditions) $\rightarrow$ lanes};
\node[box] (gate) at (1.65,-6.25)
  {\textbf{Review gate}\\the project's own bar; never edits the \textsc{bone}};

\node[box=3.3cm, draw=osAccent, line width=0.9pt] (human) at (15.75,0)
  {\textbf{\color{osAccent}Human owner}\\accountable; disposes of what the project holds true};
\node[box=3.3cm] (cur) at (15.75,-2.95)
  {\textbf{Curation}\\audits the \textsc{bone}; proposes add, update, remove, merge, split};
\node[box=3.3cm, anchor=north] (decay) at ([yshift=2.5mm]man.center -| 15.75,0)
  {\textbf{Session-start decay check}\\provenance against a sealed baseline; reports the CDI; silent when healthy};

\draw[flow] (bone.west |- agent) to node[lab, above] {read} (agent.east);
\draw[flow] (bone.east |- agent) -| node[lab, above, pos=0.35] {read} ([xshift=-9mm]human.south);

\draw[flow] (agent) to node[lab, right] {failures, blocks} (cap);
\draw[flow] (cap) to node[lab, above] {nominates} (t0);
\draw[approve] (human) to node[alab, above] {admits a rule} (t0);

\draw[flow] (agent) to node[lab, right] {plans} (oss);
\draw[flow] ([xshift=-6mm]oss.south) to node[lab, right] {change} ([xshift=-6mm]gate.north);
\draw[check] (bone.west |- man) -| node[lab, right, pos=0.75] {checked against\\decisions, contracts} ([xshift=12mm]gate.north);
\draw[flow] (gate) to node[lab, above] {merge on pass} (t2);

\draw[flow] (t2.east) -| node[lab, above, pos=0.3] {watched sources change} (decay.south);
\draw[flow] (decay.west |- man) to node[lab, above] {flags\\decayed units} (bone.east |- man);
\draw[flow] (decay) to node[lab, right] {decayed units} (cur);
\draw[approve] ([xshift=9mm]human.south) to node[alab, left] {approves\\each item} ([xshift=9mm]cur.north);
\draw[flow] (cur.west) to node[lab, above] {applies\\approved items} (bone.east |- cur.west);
\draw[check] (gate.south) |- node[lab, below, pos=0.75] {invalidation routed to curation} (17.8,-7.3) |- (cur.east);

\begin{scope}[shift={(0.0,-8.0)}, font=\scriptsize]
  \draw[flow] (0,0) to (0.7,0); \node[right] at (0.75,0) {knowledge or change};
  \draw[check] (3.6,0) to (4.3,0); \node[right] at (4.35,0) {check or routing};
  \draw[approve] (6.9,0) to (7.6,0); \node[right, text=osAccent] at (7.65,0) {human approval};
\end{scope}
\end{tikzpicture}%%
}
\caption{OSTEON as described in Section~\ref{sec:osteon}. Every fact sits at one of three tiers.
The project tier is the \bone{}, a knowledge repository with four layers and a manifest. Agents and
humans both read the \bone{}. When watched sources change, the session-start decay check compares
each unit's provenance with a sealed baseline and flags decayed units for curation. Ossification
(planned, gated change) produces a change; the review gate checks it against recorded decisions
and contracts before it merges, and routes any invalidation to curation. Curation applies only the
items a human approves. Capture records what a session learns in a ledger that governs nothing, and
a rule enters the register only when a human admits it. Solid arrows carry knowledge or change,
dashed arrows mark checks and routing, and thick blue arrows mark human approval.}
\label{fig:architecture}
\end{figure*}

\begin{table*}[t]
\caption{Requirements for superintelligent software engineering, the OSTEON mechanism for each, and
what stays with a human.}
\label{tab:mapping}
\centering
\footnotesize
\setlength{\extrarowheight}{1.5pt}
\begin{tabularx}{\textwidth}{@{}l Y Y Y@{}}
\toprule
Requirement & OSTEON mechanism & What it is built to prevent & What stays human \\
\midrule
R1 One owner per fact & Three tiers (T0 org, T1 project, T2 repository); personal tier outside; reference, never copy & A copied fact decaying in one place and not the other & Placing a contested fact \\
R2 Decisions are records & Numbered decision records; proposed, accepted, superseded; amend, never rewrite & Re-litigated or silently suspended decisions & Accepting or superseding a decision \\
R3 Seams have contracts & Versioned contracts, registry index, same-wave rule & One side of a seam changing alone & Approving a breaking version \\
R4 Gates from the project & Review against the \bone{} layers; invalidation routed to curation & Changes that pass generic review and break a project decision & Approving each change to the knowledge \\
R5 Decay checked at start & Block-level provenance; sealed baseline; Context Decay Index at session start & Work built on a fact whose source has moved & Re-verifying a decayed claim \\
R6 Planned, gated change & Daily cycle; ten-condition start gate; lanes in one session; merge is not completion & Work started on open questions; unproven completion & Answering open decisions \\
R7 No self-legislation & Capture ledger that governs nothing; nomination on corroboration & A session writing rules for the next & Admitting a rule to the register \\
\bottomrule
\end{tabularx}
\end{table*}

\subsection{Tiers and the BONE}

OSTEON places every fact at one of three tiers (R1). The organisation tier (T0) holds the
methodology: how to record a decision, version a contract, run curation. The project tier (T1)
holds the project's facts, decisions, contracts and procedures. The repository tier (T2) holds
conventions that are true inside one code repository only, in that repository's own agent
instruction file. A fact lives at the narrowest tier that fully owns it. When it could sit at two,
it is placed by what would change it, not by its topic. For example, the port a service listens
on belongs to the repository whose configuration sets it (T2), while the decision to run that
service at all belongs to the project (T1), because different events change each. Personal
preferences and an individual's memory sit outside all tiers, and the project's knowledge never
governs or cites them.

The project tier is a dedicated repository, the project's \bone{}. It holds no product code. It
holds four layers. L1 is a constitution: the project's non-negotiables, the rules the other layers
may not override, and what may not be decided unilaterally. L2 holds decisions and dated records of
the running system's state as last verified. L3 holds contracts and domain knowledge. L4 holds the
project's own skills: the procedures an agent follows inside the project. A manifest declares which
files are knowledge units and which sources are watched. Because the \bone{} is a repository, its
knowledge has what code already has: history, branches, review and provenance. The names follow one
metaphor: the project's knowledge repository is its \bone{}, knowledge that has been verified and
gated is \emph{ossified}, and work proceeds by planned ossification
(Section~\ref{sec:osteon-ossification}).

\subsection{Decision records}

Decision records follow the architecture decision record tradition~\cite{nygard2011adr}, with a
numbered identity and a lifecycle of proposed, accepted and superseded (R2). An accepted record is
amended or superseded, never rewritten, so the reason a project once had survives the decision
that replaced it. The decision skill also handles the case that matters most for agents, a merged
change that silently suspends a decision or an accepted record that no longer matches what the
running system does: the mismatch is made explicit as an amendment or a supersession, rather than
left as a contradiction between the record and the code.

\subsection{Module contracts}

Every seam between two modules has a contract (R3): the wire format, fields, units, authentication
and failure behaviour that both sides implement against. Contracts follow semantic
versioning~\cite{semver} in the spirit of design by contract~\cite{meyer1992contract}, and are
indexed in a registry. One rule carries most of the weight: a contract change ships in the
\emph{same wave}, that is, in the same change request, as the interface change it describes, never
after it. \op{} The skill also triggers when
one side of a seam is being built against the other side's source code instead of its written
contract, which is the usual way an agent learns an interface it should have read. \op{}

\subsection{Review gates and curation}

The review skill checks a change against the project's own bar, derived from its \bone{} layers,
rather than against generic practice (R4), before merge and as the self-review before a change is
declared ready. Its second job is detection. When a change invalidates a decision, a
contract or a claim in a unit, the review routes that to curation instead of passing or failing
the change on its own authority. The review skill never edits the knowledge base. One of its checks
bears directly on Eq.~\eqref{eq:escape}: a change that assumes or contradicts a fact whose state
record is past the project's staleness horizon is held as needs-information until the record is
re-verified. In release 0.1.0 that check reads the record's age, not the decay check's
source-change flag; joining the two is design work, not a result.

Curation is a separate skill that audits the \bone{} and proposes additions, updates, removals,
merges and splits of knowledge units. It applies nothing without per-item human approval. This is
where the authority axis is enforced: the agent does the reading, the comparison and the drafting;
a human approves each change to what the project holds true.

\subsection{The session-start decay check}

Every \bone{} carries a decay check wired to the agent's session-start event (R5). A knowledge unit
is a block of Markdown (a paragraph, a list item, a table row), not a file, so one changed source
decays only the blocks that cite it. A unit is \emph{anchored} when it carries at least one
resolvable provenance token naming a source: a decision record, a dated state record, a contract, a
domain document, the constitution, or a file in another declared repository. Against a sealed
baseline (the recorded digests of every watched source at the moment of the last applied curation),
a unit is \emph{decayed} when at least one of its sources has changed. The check reports a
Context Decay Index,
\begin{equation}
\mathrm{CDI} = \frac{\text{decayed anchored units}}{\text{anchored units}},
\label{eq:cdi}
\end{equation}
which is undefined, and is reported as such, when no unit is anchored. The index counts exposure,
not confirmed error: a decayed unit may still be true, and curation decides. Under the assumption
that a session consults anchored units uniformly, CDI is the project's running estimate of $s$ in
Eq.~\eqref{eq:escape}; Q3 of Section~\ref{sec:agenda} is the measurement of that quantity over
time.

Three design choices matter for the scaling argument. First, the check is silent on the healthy
path, so its cost to a session is near zero when nothing has decayed. \op{} Second, it never prints a
number for an unmeasured base: a fresh project reports the state \emph{uninitialised}, not a
reassuring zero. Third, it records every run, clean ones included, deduplicated by state (any
number of sessions in one unchanged state add one record), so the series has a denominator: the
record can answer what fraction of decay warnings led to curation, not only how many there were. Provenance and tier lints, and an attestation lint that
checks that a change to governed knowledge carries an approval record, run alongside it with a
self-test.

\subsection{Ossification: planned and gated change}
\label{sec:osteon-ossification}

Change in OSTEON proceeds by ossification (R6) in three steps. A planner skill keeps one dated cycle
per day in the \bone{} and sets the bar that every task's definition of done must clear. A starter
skill then checks the plan against ten gate conditions before any work begins; where a condition
fails on an open decision, it puts that decision to the owner as one question answerable in one
line, rather than reporting that questions exist. Only a plan that passes the gate is executed.
Execution may split the plan into independent lanes worked in parallel, by one session with
subagents or by several, and finished work is carried back into the plan: a merge is not a
completion until the plan says so. A delivery skill reports what a cycle delivered to people outside
it, with each claim graded by what proves it.

\subsection{Capture without self-legislation}

The capture skill records what a session learned about the way of working: when something fails,
when something is blocked before it lands, or when a check passes on an input that should have
failed (R7). It appends to a ledger that governs nothing. At session end it tests whether an
observation has independent corroboration; if it does, it nominates the observation and stops. A
nomination becomes a rule only through the framework's own register of named methodologies, under
human approval. Sessions learn; they do not legislate.

Two further skills keep the knowledge usable across gaps: an onboarding skill with a fixed read
order (constitution, manifest, decision and state indexes, then only the relevant skills), which
ends in an orientation whose first action names the fact it depends on; and a standup skill that
keeps a per-track continuity journal so that a dormant project can be restarted, and routes blind
spots to the role that owns them.

\subsection{What OSTEON's own history shows}
\label{sec:osteon-history}

As a check rather than as evidence, we read OSTEON's own version history. Its development spans 6
repositories and 58 commits by one author over 62 days, on 9 of which commits were made, and holds %%
54 governed artifacts (files of a governed-knowledge kind). The capture ledger received 28 %%
observations, all on three consecutive active days, which produced 3 nominations and 1 curation %%
record; no nomination can be traced by identifier into a rule, the repositories hold no decision %%
records, and capture stopped 33 days before the last commit. The same reading surfaced a likely %%
defect: 15 of the 28 ledger rows carry timestamps later than the commits that added them, which %%
suggests local time recorded under a UTC marker. Capture, nomination and one curation record were
exercised, over three days; decision records were never used; no nomination became a rule; and the
six repositories include no \bone{} of OSTEON's own, so the tiers, the gates and the decay check
were not applied to the discipline's own development. The history shows the mechanisms in partial
use by their designer and nothing more, and we report it for that reason. The collector and its
output accompany the measurement data as ancillary files.

\section{How Governance Scales in OSTEON}
\label{sec:scaling}

Table~\ref{tab:mapping} maps each requirement to its mechanism. This section relates the mechanisms
to the scaling axes and then shows three constructed scenarios.

\emph{Cost.} The decay check, the lints and the review against recorded knowledge are mechanical,
and the check is silent when nothing has decayed, so their cost per change does not grow with the
rate of change. \op{} Human effort is spent on decisions, contract versions, curation items and plan
questions, which we expect to grow far more slowly than the number of changes, because a project's
decisions are fewer and more stable than its edits. \op{} Curation
items are the exception. A decay warning is raised by a change to a watched source, and watched
sources change at a rate that tracks the change rate itself, so warnings may grow with $r$ even
when decisions do not. Whether provenance at block granularity keeps warnings per change falling is
the point on which the cost axis stands or falls, and it is Q2 of the agenda.

\emph{Coverage.} The tiers and the contract rule extend governance to every seam that has a
contract, and the contract skill is built to notice a seam that lacks one when one side is built
against the other side's source. The onboarding read order makes the governed set the first thing an agent reads, not an optional
file it may skip. \op{}

\emph{Currency.} Block-level provenance turns exposure to source change from an unobservable
property into a computed one; whether an exposed unit is in fact stale is settled in curation. In
the terms of Eq.~\eqref{eq:escape}, the decay check is built to lower the effective $s$ by flagging
decayed units before a session consults them. The review gate is built to check every change
($\rho=1$) for contradiction with the record, with its own probability $g$ in place of a reader's
$d$; it detects nothing the record does not hold current, so its value rests on the decay check
having run.

\emph{Authority.} Curation applies nothing without per-item approval, the start gate turns open
decisions into one-line questions, and the decision lifecycle forbids rewriting. The agent does
more of the work as it becomes more capable; the human disposes of the same things, at lower cost
per item. \op{}

\emph{Learning.} Capture makes the process improvable from evidence while keeping rule-making
behind corroboration and approval. \op{}

\subsection{Constructed scenarios}

The following are constructed to show mechanisms. They are not observations.

\emph{A seam drifts.} An agent changes the unit of a field one service emits, from seconds to
milliseconds, and updates its own tests, which pass. The consumer is untouched. Without a contract,
the drift surfaces in production. In OSTEON the field's unit is a clause of a versioned contract.
The same-wave rule requires the contract change in the same change request, the version bump marks
it as breaking, and the review gate flags the consumer side as built against the old version.

\emph{A decision is silently suspended.} A project once rejected client-side caching of a price, for
a reason recorded in a decision. Months later an agent, asked to cut latency, adds exactly that
cache; the change is fast, tested and well argued. Review against generic practice would approve
it. Review against the \bone{} finds the accepted decision, reports the contradiction, and routes
it. Either the change is withdrawn, or the decision is superseded by a new record that states why
the old reason no longer holds. Either way, the project's knowledge and its code agree afterwards.

\emph{A state record goes stale.} A record states that a migration has finished in production; it
cites the deployment configuration as its source. The configuration later changes. At the next
session start the decay check reports the record's blocks as decayed, before any agent builds on
the claim. Curation re-verifies the record or retires it.

In each case the agent did what it was asked well; what prevents the error is governance that
acts on the knowledge the change depended on. \op{}

\section{Design Rationale from Practice}
\label{sec:lessons}

This section states the heuristics from the author's own agent-assisted work that OSTEON was built
to (Table~\ref{tab:lessons}). They are design rationale, not findings, and the paper offers no
evidence for them. We describe no project, report no data from one, and give no episode. Each
heuristic is marked \exper{}; like \op{}, the marker flags a claim the paper does not measure.
The right column of the table says where the mechanism built to each heuristic can be checked (the
escape model of Eq.~\eqref{eq:escape}, the constructed scenarios of Section~\ref{sec:scaling}, or
OSTEON's design, Section~\ref{sec:osteon}); it does not turn the heuristic into evidence, and a
mechanism designed from a heuristic does not corroborate it. None of the heuristics is about agent
capability. Each concerns the knowledge a change relies on, or the check that guards it, which is
the object that Section~\ref{sec:argument} argues governance must act on.

\begin{table*}[t]
\caption{Design heuristics from the author's practice, each marked \exper{}, and the OSTEON
mechanism built to each. The right column says where the paper describes the mechanism; it is not
evidence for the heuristic.}
\label{tab:lessons}
\centering
\footnotesize
\setlength{\extrarowheight}{1.2pt}
\begin{tabularx}{\textwidth}{@{}>{\raggedright\arraybackslash}p{3.1cm} Y Y@{}}
\toprule
Heuristic & Statement & Built into, and where the paper describes it \\
\midrule
1. A correction is a propagation problem \exper{} &
A correction reaches the artifact that owns a fact more reliably than the artifacts that repeat it.
Writing the lesson down does not stop the same failure from coming back; a mechanical check does. &
R1, under which a fact has one owner and every other place references it; Eq.~\eqref{eq:escape},
where a mechanical gate sets $\rho=1$ and supplies its own detection $g$ in place of $d$. \\
2. A gate decays too \exper{} &
A gate is an artifact and goes stale like any other. A stale gate is worse than no gate: it still
runs, no longer tests what it claims, and turns a deliberate check into a false confirmation. &
The review factor of Eq.~\eqref{eq:escape}, whose detection term is only as good as the gate that
supplies it (the model has no term for a gate that has itself decayed); the capture skill, which
records a check that passes on an input that should fail; the self-test shipped with the lints. \\
3. Mirrors of external state decay first \exper{} &
Knowledge that mirrors external, mutable state goes stale fastest. Its staleness is driven by
events, not by elapsed time, so a review on a calendar misses it. &
Eq.~\eqref{eq:stale}, where $s$ is defined by a change to the source, not by age; the decay check,
which flags a unit when its source moves; the stale state record scenario. \\
4. Errors at a high tier are amplified \exper{} &
An error in high-tier knowledge is not repeated once; every dependent inherits it. Repair has to
happen at the source, because repairing the dependents one by one leaves the source free to seed
the error again. &
The tiers, with reference instead of copy; block-level provenance, under which one changed source
decays every block that cites it; the escape model, where one stale unit makes every change that
consults it stale-built. \\
5. Versioned references carry retractions \exper{} &
When a dependency between artifacts is a first-class, versioned reference, an assertion and its
retraction both travel along it quickly and visibly, even across repositories. &
The drifting seam scenario, where the version bump marks the change as breaking and the consumer
is flagged; provenance tokens, which may name a file in another declared repository. \\
6. A decision ledger can run at experiment speed \exper{} &
Decision records need not be slow. A ledger can keep pace with experiments and record what they
refute, provided an amendment is cheap and is recorded apart from the decision it changes. &
R2 and the amend-or-supersede lifecycle, under which the reason a project once had survives the
decision that replaced it. \\
7. A registry measures only events \exper{} &
A registry can measure only what it represents as a discrete event. A lifecycle kept as a mutable
field is overwritten as it moves, so its history cannot be measured afterwards. &
The decay check, which records every run as an event, clean runs included, so that its series has
a denominator; supersession, which is a new record rather than an edited field; the agenda's
condition on decay series (Section~\ref{sec:agenda}). \\
8. An approval that approves everything is ceremony \exper{} &
A review step that never refuses is not a filter. For review to mean anything, its refutations must
be recorded, so that what it stopped can be seen. &
Curation, which applies nothing without per-item approval; capture, which records a change blocked
before it lands; the decay record, which makes the fraction of warnings that led to curation
answerable. \\
\bottomrule
\end{tabularx}
\end{table*}

\section{Objections}
\label{sec:objections}

A thesis about the limits of capability invites five objections. We state each in its strongest
form.

\emph{``A capable enough agent will keep the documentation current itself.''} It can, but only if
it can tell which documentation is stale, which requires knowing which source each claim came from
and whether that source has changed since. That is provenance, and it is governance. An agent that
rewrites documentation without it produces documentation that agrees with the code, not
documentation that agrees with what the project decided. \op{} The objection is an argument for R5, not
against it.

\emph{``The tests are the specification.''} Tests specify behaviour at the points they exercise.
They do not record why a behaviour was chosen, which alternatives were rejected, who owns a field
on the far side of a seam, or what is running in production. A change can pass every test and
reverse a decision. The training route that uses test patches as specifications~\cite{wei2026ssr}
is right to do so for learning to repair code; a project needs more than that to be governed.

\emph{``Governance will slow the agents down and teams will turn it off.''} This is the strongest
objection, and it is why cost is the first scaling axis. Governance that costs a human minutes per
change will be skipped once changes arrive faster than humans can spend those minutes. The answer
is governance whose per-change cost is mechanical and whose human cost attaches to decisions.
Whether OSTEON achieves that is Q2 of the agenda; we do not claim it has been shown.

\emph{``This is decision records and documentation as code, renamed.''} The mechanisms have long
roots, and we cite them in Section~\ref{sec:related}. Three things are new. Decay is computed per
Markdown block, for knowledge that lives outside the code (decisions, contracts, state records),
from declared provenance to its sources against a sealed baseline, at every session start, rather
than by comparing text with code after the fact or when a link breaks; inconsistency detection
between code and its comments already does this for knowledge inside the
code~\cite{panthaplackel2021jit}, and we extend the idea to knowledge the code does not carry. The
human authority boundary is stated as an invariant that does not move with capability. And the
whole is defined as a discipline for a regime in which the reviewer can no longer verify by
reading, which the older practices did not have to address.

\emph{``Let agents review agents.''} An agent reviewer raises $c$. If it reviews by reading, it
inherits the reviewer's problem at a higher rate: it judges the change against knowledge it cannot
know is current. If it reviews against recorded decisions and contracts, it is a gate, and its
detection is bounded by the currency of the record (Section~\ref{sec:arg-throughput}). Either way
the knowledge, not the reviewer, is the term that governance has to act on.

\section{Related Work}
\label{sec:related}

\begin{table*}[t]
\caption{Positioning against related work.}
\label{tab:related}
\centering
\footnotesize
\setlength{\extrarowheight}{1.5pt}
\begin{tabularx}{\textwidth}{@{}>{\raggedright\arraybackslash}p{3.0cm} Y Y@{}}
\toprule
Work & Primary object, and the human's role in it & Relation to this paper \\
\midrule
SE 3.0~\cite{hassan2024se3} & AI-native, intent-centric development stack; the developer works in partnership with AI teammates & Names the era; we name the discipline for the regime past delegation \\
Agentic SE~\cite{hassan2025agentic} & SE for Humans and SE for Agents, with command and execution environments; humans orchestrate and mentor agent teams & Closest vision; we make governance scaling the defining property and supply a released discipline \\
Self-play SWE-RL~\cite{wei2026ssr} & Agent training by self-play bug injection and repair; humans supply sandboxed repositories & The capability route; we argue it does not act on knowledge decay \\
Programming with trust~\cite{roychoudhury2026trust} & Trust in agent-written code via analysis evidence; humans set the trust bar & Same premise; we attach verification to recorded decisions and contracts rather than to the change \\
Evaluating AGENTS.md~\cite{gloaguen2026agentsmd} & Effect of context files on agent task success; humans write the files & Read by us as evidence that more ungoverned context is not the lever \\
Context-file studies~\cite{chatlatanagulchai2025manifests,mohsenimofidi2025context,chatlatanagulchai2025readmes,lulla2026agentsmd} & Content, adoption and evolution of context files, and their effect on agent efficiency; humans write and maintain the files & We check whether the paths a file names resolve in its repository's tree \\
ContextNest~\cite{sulpovar2026contextnest} & Governed knowledge vaults beneath retrieval; humans approve what agents may consume & Complementary integrity and provenance layer; we govern the engineering lifecycle \\
Deliberative Curation~\cite{johnson2026deliberative} & Multi-agent curation protocol for shared knowledge; humans design the protocol & Agents curate collectively; OSTEON keeps per-item human approval \\
\bottomrule
\end{tabularx}
\end{table*}

Table~\ref{tab:related} summarises the positioning.

\emph{SE 3.0 and agentic software engineering.} Hassan et al.\ propose SE 3.0, an AI-native
software engineering characterised by intent-centric, conversation-oriented development with AI
teammates, and a technology stack and roadmap for it~\cite{hassan2024se3}. In later work, Hassan
and colleagues describe agentic software engineering through a duality of SE for Humans and SE for
Agents. Humans
orchestrate and mentor agent teams in an Agent Command Environment, and agents work and call on
human expertise in an Agent Execution Environment; structured artifacts such as merge-readiness
packs and consultation request packs carry the exchange~\cite{hassan2025agentic}. That vision is
the nearest to ours, and several OSTEON mechanisms read as instances of it: the start gate's
one-question-at-a-time decisions resemble consultation requests, and review against recorded
knowledge resembles an evidence-backed merge pack. We differ in two respects. We define the regime
by verification capacity, so the discipline is about what happens when the human can no longer
review by reading; and we make the scaling of governance, rather than the partnership, the defining
property, and accompany the argument with a released artifact.

\emph{Programming with trust.} Roychoudhury et al.\ frame the shift from programming at scale to
programming with trust, and propose that trust in agent-written code be built from analysis
evidence that travels with the change~\cite{roychoudhury2026trust}. We share the premise that
review by reading will not carry the load. We differ in where the evidence attaches: their trust
artifacts certify the change against the code; our gates certify it against the project's recorded
decisions and contracts, which the code does not carry.

\emph{The training route.} Self-play SWE-RL trains an agent on real repositories with minimal data
assumptions, specifies each bug by a test patch, and reports self-improvement on SWE-bench Verified
and SWE-Bench Pro~\cite{wei2026ssr}. We see it as the capability half of the regime,
complementary to ours: such training does not see decisions, contracts or the current state of a
running system, that knowledge decays independently of the agent, and only governance acts on it.

\emph{Context files.} Gloaguen et al.\ evaluate repository-level context files in a controlled
study. They find that the files do not generally improve task success while raising cost, that
instructions are followed, and that overviews do not help~\cite{gloaguen2026agentsmd}. In a
smaller study of 124 pull requests in 10 repositories, Lulla et al.\ associate \texttt{AGENTS.md}
files with lower agent runtime and output token use~\cite{lulla2026agentsmd}. Earlier empirical
studies catalogue the content, adoption and evolution of context files in GitHub
repositories~\cite{chatlatanagulchai2025manifests,mohsenimofidi2025context,chatlatanagulchai2025readmes};
our measurement differs in checking, against each repository's tree, whether the paths a file
names resolve. The two evaluations disagree on cost, one finding it higher and the other lower; we
take no side, and our measurement bears on neither.
OSTEON's knowledge units are a different object from a context file: versioned,
provenance-anchored, checked for decay and owned at one tier. Whether that difference changes the
result is an open empirical question (Q4). Our measurement in Section~\ref{sec:measure} is
complementary to their evaluation: it describes how such files age in public repositories, not how
they affect agents.

\emph{Context governance.} ContextNest formalises context governance for knowledge that agents
consume, with typed documents, deterministic selectors, hash-chained version histories, checkpoints
and audit traces, so that an organisation can reconstruct which knowledge versions informed an
agent output~\cite{sulpovar2026contextnest}. It is the closest in mechanism. Its object is the
eligibility and integrity of knowledge beneath retrieval; ours is the engineering lifecycle of
project knowledge: decisions, contracts, decay against moving sources, and planned change. The two
compose; a \bone{} could sit on an integrity layer of that kind.

\emph{Collective curation.} Deliberative Curation proposes a protocol for multi-agent knowledge
bases, with a formalised lifecycle, reputation-weighted deliberative voting and graduated
sanctions, evaluated in simulation~\cite{johnson2026deliberative}. OSTEON keeps a human as the
disposing authority for every change to project knowledge. As agent counts grow, per-item human
approval may become the binding cost, and protocols of that kind may be needed to decide what
reaches the human at all; Q6 lists this as open.

\emph{Decisions, contracts and decay.} The record of design decisions as first-class architectural
knowledge~\cite{jansen2005decisions,tyree2005decisions}, the knowledge-management work that
followed~\cite{capilla2016akm}, the lightweight decision record~\cite{nygard2011adr}, design by
contract~\cite{meyer1992contract}, semantic versioning~\cite{semver}, provenance
modelling~\cite{moreau2013prov}, the laws of software evolution and ageing
\cite{lehman1980laws,parnas1994aging}, and the technical debt metaphor under which stale
documentation is a debt that accrues interest~\cite{kruchten2012debt} are the foundations OSTEON
builds on. What is new is not any one mechanism but their combination into a discipline that agents
operate and humans govern, with decay made computable at session start.

\emph{Stale documentation and traceability.} That documentation drifts from code is established:
comments and code co-evolve only loosely~\cite{fluri2007coevolve}, and inconsistencies between
them have been detected by rule~\cite{tan2007icomment}, by identifier
analysis~\cite{ratol2017fragile}, at scale~\cite{wen2019inconsistencies} and at change time, before
anything breaks~\cite{panthaplackel2021jit}; outdated content is among the most reported
documentation problems~\cite{aghajani2019docissues}, and the maintenance of trace links between
artifacts is a research field of its own~\cite{clelandhuang2014traceability}. The decay check
differs in its object and its oracle: it watches knowledge that the code does not carry, and it
compares declared provenance against a sealed baseline of its sources rather than text against
code. The link check of Section~\ref{sec:measure} is the routine kind. Its closest relatives are DOCER,
which flags code elements that documentation still names after every instance has left the source
and finds outdated references in 28.9\% of the 918 projects it analysed~\cite{tan2024docer}, and the
unmodified use of DOCER on agent configuration files by Treude and Baltes, who call the result
context rot and find 1.27\% of 18,048 verified references stale, with at least one in 23.0\% of 356
repositories~\cite{treude2026contextrot}. Our link check is coarser than both: it reads file paths
against a repository's tree at one commit rather than code element names against the source, and it
reports a level, not a rate.

\emph{Scalable oversight.} Work on AI safety frames the problem of supervising systems that exceed
their supervisors~\cite{amodei2016concrete,irving2018debate,bowman2022scalable,burns2023weaktostrong}.
Superintelligent software
engineering is a domain instance of that problem, and our move is domain-specific: change the
object of oversight from the change to the governed knowledge it is checked against.

\section{Research Agenda}
\label{sec:agenda}

A vision earns its keep by saying what would prove it wrong. We state the thesis as testable
claims, with the evidence that would refute each, and close with one design question (Q6).

\emph{Q1: Does governed knowledge reduce stale-built change?} Fix a set of decisions and contracts
for a task set, written by the experimenters. Expose them to the agent as governed knowledge in one
arm and as ordinary documentation, or not at all, in the other, at matched agent capability, and
measure the fraction of merged changes that contradict them; the experimenters' set is the oracle,
so it is defined in both arms. Vary throughput at fixed review capacity within each arm. The thesis
is weakened if the fraction does not differ between arms, if capability alone drives it to near
zero, or if it does not rise with throughput once reviewers are saturated.

\emph{Q2: Does governance cost scale sub-linearly in throughput?} Measure total human minutes spent
on decisions, contract versions, curation items and plan questions against the number of changes,
in the same projects as Q1 so that the effect is held while the cost is read. The scaling principle
fails if human governance effort grows in proportion to changes, or falls only because governance
is skipped.

\emph{Q3: How fast does project knowledge decay?} Measure time to decay of anchored units against
their sources, and the fraction of decay warnings that lead to curation. A decay rate near zero
would make R5 unnecessary. Section~\ref{sec:measure} gives a coarse, cross-sectional reference
point, not a rate, for ungoverned agent context files; the rate for governed knowledge remains
open.

\emph{Q4: Do context results transfer?} Repeat the context-file evaluation of Gloaguen et
al.~\cite{gloaguen2026agentsmd} with tasks whose correct fix depends on a recorded decision not
visible in the code, in two conditions: the decision is current, and the decision has been
superseded since the file was written. In the first condition a context file and a governed unit
present the same text; in the second only the governed unit is flagged as decayed. The claim that
governed knowledge differs from context files fails if the two arms do not differ in the second
condition.

\emph{Q5: Do gates hold against a more capable agent?} Test whether review against recorded
knowledge still detects contradictions when the agent argues persuasively for its change (the
setting studied as debate~\cite{irving2018debate}), and whether detection degrades as agent
capability rises. The gate claim fails if detection falls as capability rises, because $g$ is then
a property of the agent after all and the escape rate regains its dependence on the agent.

\emph{Q6: When may agents curate?} Compare per-item human approval with deliberative multi-agent
curation~\cite{johnson2026deliberative} on error rate and human cost, to find where the authority
boundary can be priced lower without moving. This is an open design question rather than a test of
the thesis; it bounds how far the cost of the authority axis can fall without moving it.

Two conditions apply to all six: instruments should be released even when the data cannot be,
because the data of real projects often cannot leave them; and decay series should include clean
runs, so that rates have denominators, which is why OSTEON's decay check records them.

\emph{Registration.} A controlled evaluation of Q1, Q4 and Q5 is in preparation; its task
protocol and hypotheses will be self-registered on the Open Science Framework before any
confirmatory run. Two registrations are public: a commitment to the development seed,
\url{https://osf.io/s4w2t}, and the SHA-256 of the sealed list of candidate pairs from a census
of public architecture decision records, \url{https://osf.io/8scgm}.

\section{Limitations}
\label{sec:limits}

The paper reports no measurement of OSTEON's effect, so nothing here shows that the discipline
works; it shows what it does and why we expect it to matter. The measurement in
Section~\ref{sec:measure} is a single observational snapshot of popular public repositories. It
counts unresolved path references in agent context files, a possible symptom of decay, not wrong
instructions or their effect on agents, and its headline rate depends strongly on how references
are read (Sections~\ref{sec:measure-sens} and~\ref{sec:measure-threats}). The
model in Section~\ref{sec:argument} assumes independent staleness, a fixed review capacity and a
fixed detection probability. Real projects violate all three. The model also holds $s$ fixed as $r$
rises. The agent's own changes are among the sources that make other units stale, so in practice
$s$ rises with $r$, and the model understates the effect in that respect. The model is meant to
show the direction of the effect, not its size. The definition rests on a threshold relative to a
project's own engineers that mixes capability with throughput, is not crisp, and will be crossed at
different times for different tasks; a reader who prefers to name the regime by its operative
condition, verification capacity, loses nothing in the argument.
OSTEON 0.1.0 targets one agent platform, and its mechanisms are designed by one author; independent
use may expose costs we have not seen. Finally, governance has a cost, and a team may reasonably
refuse it while agents still work in the delegated regime. Our claim concerns the superseded
regime, where we argue the alternative is worse. What OSTEON's own history shows, and does not
show, is in Section~\ref{sec:osteon-history}.

\section{Conclusion}

Training is making software agents more capable, and it may make them more capable than the
engineers who answer for the systems they change. We have argued that this capability is not enough, for three reasons. First, correctness is relative
to decisions and contracts that the code does not carry. Second, this knowledge decays. Third, once
agents outpace their reviewers, the changes built on stale knowledge escape review at a rate that
grows with throughput. A snapshot of popular public repositories shows that the mismatch the second
reason predicts is common, and does not show that it accumulates. Agent context files are common,
and many of them stand still through hundreds of commits. Most of those that name repository paths
name at least one that does not resolve. Files that stood still through more commits, however, do
not have more unresolved references. Superintelligent software engineering is the discipline that
answers this by scaling governance with capability. OSTEON is one concrete discipline for it,
released with its scripts and data~\cite{osteon2026zenodo}. Whether it works is an empirical
question, and we have stated the evidence that would settle it.

\section*{Availability}

OSTEON 0.1.0, including the fifteen skills, the knowledge-repository scaffold and the decay check,
is archived on Zenodo with DOI \mbox{10.5281/zenodo.23436020}~\cite{osteon2026zenodo}. Content is licensed
under CC BY 4.0 and scripts under MIT.

The measurement in Section~\ref{sec:measure} is provided as ancillary files with this preprint:
the pre-specified protocol with its dated deviation note (\texttt{README.md}); the collection,
summary, figure, post hoc audit and revision scripts, in Python~3 (\texttt{collect.py},
\texttt{summarize.py}, \texttt{figures.py}, \texttt{audit.py}, \texttt{extend.py}; all but the
figure script use the standard library only); the dataset (\texttt{frame.csv}, one row per
repository returned by the search; \texttt{repos.csv}, one row per context file, with a
symbolic-link flag; \texttt{funnel.json}; \texttt{summary.json}; \texttt{candidates.csv};
\texttt{audit.json}; and \texttt{extended.json}, the statistics added at revision); and the model
behind Fig.~\ref{fig:escape} (\texttt{escape\_model.py}, \texttt{fig-escape-model.json}).

\section*{Acknowledgment}

The author used an AI assistant in drafting and editing this paper, reviewed the text, and takes
full responsibility for its content.

\IEEEtriggeratref{25}

\end{document}